\documentstyle[12pt,epsf]{article}
\begin{document}
\parskip=0pt
\parindent=123mm

{\hskip -10mm} TECHNION-PH-96-21\par
{\hskip -10mm} DESY-97-051\par

\bigskip
\bigskip
\begin{center}
\parskip=10pt

{\Large\bf   Calculations of penguin diagrams in B decays}
\bigskip
\bigskip

{\large   Cai-Dian L\"u$^a$\footnote{Alexander von Humboldt fellow.} and Da-Xin Zhang$^b$ }

$a$ II Institut f\"ur Theoretische Physik, Universit\"at Hamburg,
D - 22761 Hamburg, Germany

$b$ Physics Department, Technion- Israel Institute of Technology,
 Haifa 32000, Israel

\end{center}

\parindent=23pt
\bigskip
\bigskip
\bigskip
\bigskip
\bigskip
\bigskip
\begin{abstract}
We analyze the effects of the space-like penguin
diagrams in two body B decays $B^-\to K^-{\bar K}^0$
and $B^-\to \pi^-{\bar K}^0$.
Special attention is paid to the operator $Q_6$,
whose contribution is large in the BSW model,
but vanishes in the PQCD method in the approximation
of neglecting the masses in the final states.
Suppressions of the space-like penguin
diagrams are found in the PQCD method,
which implies that the contributions of the space-like penguins
are small compared to those of the time-like penguins.
\end{abstract}

\bigskip
\bigskip

\noindent {\it PACS number(s)}: 13.25Hw, 12.15Mm, 14.40Nd, 12.38Bx
\vfill
\newpage

\section{} 

Penguin diagrams in exclusive B decays have been of great interests
in the study of CP violations\cite{1}.
Up to now, most of the quantitative investigations 
concentrate on the so-called time-like penguin diagrams,
except in \cite{xing,yang} where, 
using  the BSW model\cite{bsw}, 
the space-like penguin diagrams are found to be important.

An important feature of the  BSW model in nonleptonic decays
is the use of the factorization hypothesis\cite{bsw}.
Under this hypothesis,
one decomposes the decay amplitude into the
product of two hadronic matrix elements which are related
to the  leptonic or semileptonic processes.
While this hypothesis works  for the tree level processes
if some phenomenological parameters are introduced to
fit the experimental data\cite{bsw,pheno},
there exists no firm foundation to use it in processes involving
penguin diagrams.
Because the penguin operators have different Lorentz
structures from the tree level $(V-A)\times (V-A)$ ones,
it is necessary to use the equations of motion for the free
quarks to relate the hadronic matrix elements
of these two kinds of operators\cite{eom}.
However, this procedure is not proved
since these equations must be used for the quarks
inside hadrons.
Consequently, for a special operator $Q_6$\cite{buras},
the decay amplitudes of the space-like penguins always 
involve the  factor $M_B/(m_q-m_{q'})$,
where $q$ and $q'$ are light quarks\cite{eom,xing,yang}.
This  is a huge factor and more seriously, 
it might produce the singularity
at $m_q=m_{q'}$ which cannot be cured by the BSW model itself.

In this work we use the perturbative QCD(PQCD) method\cite{pqcd,wyler}
to re-analyze the penguin dominated processes $B^-\to K^-{\bar K}^0$
and $B^-\to \pi^-{\bar K}^0$.
The calculations of the decay amplitudes between the involved hadrons
are replaced by those of the scattering amplitudes between the quarks,
and the non-perturbative aspects of QCD are attributed to
the wave functions which are taken as external inputs.
Since there  exists no reliable way to determine these  wave functions
up to now,
some approximate forms (like those to be used below) were chosen in the 
literature as practical ways of performing phenomenological analysis.
Hence the results  are model dependent,
like those from the BSW model\cite{bsw}.

Following \cite{wyler}, in the PQCD method
there is no need of using the factorization hypothesis or the 
equation of motion in the nonleptonic processes.
We calculate the decay amplitudes for the penguin diagrams,
and compare the contributions from the
space-like and time-like penguins.
We find that within PQCD there is no singularity which exists 
in the BSW model,
and the space-like penguins are generally suppressed 
compared to the time-like ones.

\section{} 

We take the pure penguin processes as the examples in the
following.
To start,
we use the effective Hamiltonian for the charmless  nonleptonic B decays
 as
\cite{buras}
\begin{equation}
{\cal H}_{eff}=\displaystyle\frac{G_F}{\sqrt{2}}\sum_{i=1}^6 V_{CKM} C_i(m_b)Q_i,
\label{heff}
\end{equation}
where $V_{CKM}$ denotes the CKM factors, 
and the operators are
\begin{equation}
\begin{array}{rlrl}
Q_1^u= & (\bar{q}_{\alpha}u_{\beta})_{V-A}
      (\bar{u}_{\beta}b_{\alpha})_{V-A}, &
Q_2^u= & (\bar{q}u)_{V-A}(\bar{u}b)_{V-A}, \\
Q_3= & (\bar{q}b)_{V-A}\displaystyle\sum_{q'}(\bar{q}'q')_{V-A},  &
Q_4= & (\bar{q}_{\alpha}b_{\beta})_{V-A}
      \displaystyle\sum_{q'}(\bar{q}'_{\beta}q'_{\alpha})_{V-A},  \\
Q_5= & (\bar{q}b)_{V-A}
      \displaystyle\sum_{q'}(\bar{q}'q')_{V+A},  &
Q_6= & (\bar{q}_{\alpha}b_{\beta})_{V-A}
      \displaystyle\sum_{q'}(\bar{q}'_{\beta}q'_{\alpha})_{V+A}
\end{array}
\label{oper}
\end{equation}
for $q=d, ~s$ and $q'=u, ~d, ~s$.
The operators $Q_3$ to $Q_6$ come from the penguin diagrams.
Performing the Fierz transformations,
we get 
\begin{equation}
\begin{array}{rlrl}
Q_1^u= & (\bar{u}u)_{V-A}(\bar{q}b)_{V-A}, &
Q_2^u= & (\bar{q}u)_{V-A}(\bar{u}b)_{V-A},\\
Q_3= &\displaystyle\sum_{q'} (\bar{q}b)_{V-A}(\bar{q}'q')_{V-A}, &
Q_4= &\displaystyle\sum_{q'} (\bar{q}'b)_{V-A}(\bar{q}q')_{V-A}, \\
Q_5= & \displaystyle\sum_{q'} (\bar{q}b)_{V-A}
     (\bar{q}'q')_{V+A}, &
Q_6= & -2 \displaystyle\sum_{q'}(\bar{q}'b)_{S-P}(\bar{q}q')_{S+P}.
\end{array}
\label{fierz}
\end{equation}
The Wilson coefficients of (\ref{heff}) are\cite{buras}
\begin{equation}
\begin{array}{rlrl}
C_1(m_b)=& -0.184   ,&C_2(m_b)= & 1.078  , \\
C_3(m_b)=&  0.013  ,&C_4(m_b)= & -0.035  , \\
C_5(m_b)=&  0.009  ,&C_6(m_b)= & -0.041  .\\
\end{array}
\label{wilson}
\end{equation}
Then the amplitudes for the 2-body nonleptonic B decays are
\begin{equation}
<YZ|{\cal H}_{eff}|B>=\displaystyle\frac{G_F}{\sqrt{2}}
\sum_{i=1}^6 V_{CKM}C_i(m_b)<YZ|Q_i|B>.
\label{amplitude}
\end{equation}
In processes with penguin contributions, 
the amplitudes of (\ref{amplitude})
receive contributions from both time-like and space-like penguin diagrams,
which are depicted in Figure 1 (a) and 1 (b).

\section{} 

Working in the BSW model,
the treatments of the space-like penguin contributions of the operator
$Q_6$ need to be discussed in some details.
These treatments can be described as the following.
We use the basis in  (\ref{fierz}).
First,
the factorization hypothesis is made to get
\begin{equation}
<Y(q'\bar{q}_V)Z(q_V\bar{q}')|Q_6|B(b\bar{q}')>=
-2<Y(q'\bar{q}_V)Z(q_V\bar{q}')|(\bar{q}q')_{S+P}|0>
<0|(\bar{q}'b)_{S-P}|B(b\bar{q}')>.
\label{factor}
\end{equation}
Second,
use of the equations of motion\cite{xing,yang,eom}
\begin{equation}
\bar{q} q'=\displaystyle
\frac{-i\partial^\mu(\bar{q}\gamma_\mu q')}{m_q-m_{q'}}, ~~~
\bar{q}'\gamma_5 b=\displaystyle
\frac{-i\partial^\mu(\bar{q}'\gamma_\mu\gamma_5 b)}{m_b+m_{q'}}
\label{eom}
\end{equation}
is needed to relate the two hadronic matrix elements on the right-handed-side
of (\ref{factor}) to those with the usual ones with $V\pm A$ currents.
Third,
the quark currents on the right-handed-sides of (\ref{eom})
are approximated by their hadronic counterparts as\cite{bsw}
\begin{equation}
\bar{q}b=(\bar{q}b)_H,
\label{hadron}
\end{equation}
where the sublabels denote the hadronization.
And fourth,
partial differential operations on the hadronic matrix elements 
are used giving the final result
\begin{equation}
\begin{array}{rcl}
<Y(q'\bar{q}_V)Z(q_V\bar{q}')|Q_6|B(b\bar{q}')>&=&
\displaystyle\frac{2m_B^2}{(m_q-m_{q'})(m_b+m_{q'})}\\&&
<Y(q'\bar{q}_V)Z(q_V\bar{q}')|(\bar{q}q')_{V}|0>
<0|(\bar{q}'b)_{A}|B(b\bar{q}')>.
\end{array}
\label{factor-result}
\end{equation}

There exist another  ambiguity in the procedure described above
besides the usefulness of the factorization hypothesis.
In using the equations of motion in (\ref{eom}) we have worked
at the quark level, 
thus the quark masses appearing in (\ref{eom}) seem to be the current ones.
However,
it is also allowed to reverse the second and the third steps in the above 
procedure, 
which gives no explanation of which mass should be used.
Furthermore,
when $q'$ and $q$  are the same flavors then
(\ref{factor-result}) results in infinity which violates the analytic
property of the amplitude.
Even without the singularity,
the  factor $m_B/(m_{q}-m_{q'})$ in (\ref{factor-result})
enhances the amplitude of $Q_6$ greatly, 
which is difficult to understand.

\section{} 

Now we turn to the PQCD calculations of the penguin amplitudes.
In this method,
no explicit use of the factorization hypothesis is needed.
We cannot resolve the ambiguity of the quark masses described above.
However, this ambiguity is less serious than that in the BSW model.

Following \cite{wyler},
we take the interpolating fields for the mesons as
\begin{equation}
\begin{array}{rcl}
\psi_B&=&\displaystyle\displaystyle\frac{1}{\sqrt{2}}
\displaystyle\displaystyle\frac{I_c}{\sqrt{3}}
\phi_B(x)\gamma_5({\not\! p}-m_B),\\
\psi_Y&=&\displaystyle\displaystyle\frac{1}{\sqrt{2}}
\displaystyle\displaystyle\frac{I_c}
{\sqrt{3}}\phi_Y(y)\gamma_5 \not\!p_Y,\\
\psi_Z&=&\displaystyle\displaystyle\frac{1}{\sqrt{2}}
\displaystyle\displaystyle\frac{I_c}
{\sqrt{3}}\phi_Z(z)\gamma_5 \not\! p_Z.
\end{array}
\label{wf}
\end{equation}
Here $I_c$ is  an identity in the color space, 
and we have neglected the masses of the final states.

The decay amplitudes can be calculated using Figure 2 and 3 for the
time-like and the space-like penguins, respectively.
We  neglect terms proportional
to the final state masses,
and keep in each diagram the leading term in the expansion in $x_1$,
the momentum fraction carried by the light anti-quark in the
B meson.
Denoting
\begin{equation}
<YZ|Q_i|B>=A_{ia}+A_{ib}+\cdots +A_{if},
\end{equation}
where $A_{ik}(k=a,b,\cdots f)$ is the contribution from the
$k$-th diagram of Figure 2 or 3.
When the diagrams have the topology which can induce
the processes,
the results are summarized in the following.

\noindent
(a) The time-like penguins:

\begin{eqnarray}
A_{3c}^{time} &=&-\displaystyle\frac{8g_s^2}{3\sqrt{6}}
\int_0^1 [{\rm d}x][{\rm d}y][{\rm d}z]
\phi_B(x)\phi_Y(y)\phi_Z(z)
\displaystyle\frac{4(-2x_1+y_1+1)}{x_1(x_1-y_1)(2x_1-y_1)},\nonumber\\
A_{3d}^{time}+A_{3e}^{time} &=& -\displaystyle\frac{8g_s^2}{3\sqrt{6}}
\int_0^1 [{\rm d}x][{\rm d}y][{\rm d}z] \phi_B(x)\phi_Y(y)\phi_Z(z)
\displaystyle\frac{4}{x_1(x_1-y_1)(x_1-z_1)},\label{o3time}
\\
A_{3a}^{time}&=& A_{3b}^{time}~=~A_{3f}^{time}= 0,\nonumber\\
A_{4a}^{time} &=& -\displaystyle\frac{8g_s^2}{3\sqrt{6}}
\int_0^1 [{\rm d}x][{\rm d}y][{\rm d}z] \phi_B(x)\phi_Y(y)\phi_Z(z)
\displaystyle\frac{12(-2x_1+y_1+1)}{x_1(x_1-y_1)(2x_1-y_1)},\nonumber\\
A_{4b}^{time}&=& A_{4c}^{time}~=~A_{4d}^{time} ~=~A_{4e}^{time}
 ~=~A_{4f}^{time}= 0,
\end{eqnarray}

\begin{eqnarray}
A_{5a}^{time} &=&
A_{5b}^{time} =
A_{5c}^{time} =
A_{5d}^{time}+A_{5e}^{time} =
A_{5f}^{time} = 0,\\
A_{6a}^{time} &=&
A_{6b}^{time} =
A_{6c}^{time} =
A_{6d}^{time}=A_{6e}^{time} =
A_{6f}^{time} = 0,
\label{o5time}
\end{eqnarray}
where $[{\rm d}x]={\rm d}x_1{\rm d}x_2$,  $[{\rm d}y]={\rm d}y_1{\rm d}y_2$,  
$[{\rm d}z]={\rm d}z_1{\rm d}z_2$.

\noindent
(b) The space-like penguins:

\begin{equation}
\begin{array}{rcl}
A_{3a}^{space} &=& A_{3b}^{space}=A_{3e}^{space}+A_{3f}^{space}=0,\\
A_{3c}^{space} &=& \displaystyle\frac{8g_s^2}{3\sqrt{6}}
\int_0^1 [{\rm d}x][{\rm d}y][{\rm d}z] \phi_B(x)\phi_Y(y)\phi_Z(z)
\displaystyle\frac{4}{(1-y_1)(x_1-x_1y_1+x_1z_1-y_1z_1+y_1-1)},\\
A_{3d}^{space} &=& \displaystyle\frac{8g_s^2}{3\sqrt{6}}
\int_0^1 [{\rm d}x][{\rm d}y][{\rm d}z] \phi_B(x)\phi_Y(y)\phi_Z(z)
\displaystyle\frac{4}{z_1(1-y_1)(x_1-z_1)},
\end{array}
\label{o3space}
\end{equation}

\begin{equation}
\begin{array}{rcl}
A_{4a}^{space} &=& -\displaystyle\frac{8g_s^2}{3\sqrt{6}}
\int_0^1 [{\rm d}x][{\rm d}y][{\rm d}z] \phi_B(x)\phi_Y(y)\phi_Z(z)
\displaystyle\frac{-12}{z_1(1-y_1)},\\
A_{4b}^{space} &=& -\displaystyle\frac{8g_s^2}{3\sqrt{6}}
\int_0^1 [{\rm d}x][{\rm d}y][{\rm d}z] \phi_B(x)\phi_Y(y)\phi_Z(z)
\displaystyle\frac{12}{z_1(1-y_1)},\\
A_{4k}^{space} &=&  0 ~~~~(k=c,d,e,f),
\end{array}
\label{o4space}
\end{equation}

\begin{equation}
A_{5k}^{space} =  0 ~~~~(k=a,b,\cdots f),
\label{o5space}
\end{equation}

\begin{equation}
A_{6k}^{space} =  0 ~~~~(k=a,b,\cdots f).
\label{o6space}
\end{equation}
In (\ref{o3time}-\ref{o6space}) $g_s=\sqrt{4\pi\alpha_s}$
is the coupling constant of QCD.

Independent of the special choices of the wavefunctions,
we can draw some general conclusions in the
PQCD method.
First,
from (\ref{o6space}),
the space-like penguin contribution from the operator $Q_6$ vanishes,
and thus the singularity due to the degeneration 
of the quark masses disappears.
This is in sharp contrast to the observation within the BSW model
discussed above.

Second,
comparing the right-hand-sides of equations
(\ref{o3space}-\ref{o6space}) and those of 
 equations (\ref{o3time}-\ref{o5time}) ,
it is obvious that the contributions from the space-like penguins
are always suppressed compared to those from
the time-like penguins by a factor of $x_1$,
the momentum fraction carried by the light quark
inside the B meson, 
which  is taken to be $0.05<\!<x_1<\!<0.1$\cite{pqcd,wyler} 
in the PQCD method.
This suppression of the space-like penguins
reflects the mechanism of the helicity-suppression,
as stated in \cite{eom}.

\section{}

To get the quantitative estimates,
we  take the wave functions  as \cite{pqcd,wyler}
\begin{eqnarray}
\phi_{B}(x)&=&\displaystyle\frac{f_B}{2\sqrt{3}}\delta(x_1-\epsilon_B),\\
\phi_Y(y)&=& \sqrt{3} f_Y y_1 (1-y_1),\\
\phi_Z(z)&=& \sqrt{3} f_Z z_1 (1-z_1).
\end{eqnarray}
Performing the integrations over $[{\rm d}x]$, $[{\rm d}y]$ and $[{\rm d}z]$, 
we get the final amplitudes:

\begin{equation}
\begin{array}{rcl}
A_{3c}^{time} &=& \Delta
(\displaystyle\frac{1}{2}-2\ln\frac{1-2\epsilon_B}{2\epsilon_B}
+\ln\frac{1-\epsilon_B}{\epsilon_B}-i\pi)\displaystyle\frac{1}{\epsilon_B},\\
A_{3d}^{time}+A_{3e}^{time}&=&-\displaystyle\frac{3}{2} 
\Delta\displaystyle\frac{1}{\epsilon_B},
\end{array}
\end{equation}

\begin{equation}
\begin{array}{rcl}
A_{4a}^{time} &=& 3\Delta
(\displaystyle\frac{1}{2}-2\ln\frac{1-2\epsilon_B}{2\epsilon_B}
+\ln\frac{1-\epsilon_B}{\epsilon_B}-i\pi)\displaystyle\frac{1}{\epsilon_B},\\
\end{array}
\end{equation}

\begin{equation}
\begin{array}{rcl}
A_{3c}^{space} &=& -0.936\Delta ,\\
A_{3d}^{space} &=& 3\Delta(1-\ln\frac{1-\epsilon_B}{\epsilon_B}-i\pi),
\end{array}
\end{equation}
when the diagram contributes to the specified decay process.
The contributions of all the other diagrams vanish.
Here $\Delta=\displaystyle\frac{4\sqrt{2}}{9} g_s^2f_Bf_Yf_Z$.

The branching ratios for 
$B^-\to K^-{\bar K}^0$ and $B^-\to\pi^-{\bar K}^0$, 
 are displayed in table 1.
In the numerical calculations, 
we have used $\tau_B=1.62\times 10^{-12}s$,  $f_B= 0.132$GeV, $f_K=0.113$GeV, 
$f_\pi = 0.093$GeV, $|V_{ts}^*V_{tb}|=0.044$ and  $|V_{td}^*V_{tb}|=0.013$.

 From Table 1, it can be found that the space-like penguins contribute only
several percents to the branching ratios and are thus less important.
This result is consistent with the analysis we have carried out in the
previous section.

\begin{table}
\begin{tabular}{c|cc|cc}
\hline
\hline
&  BR($B^- \to \pi ^- K ^0$)& ($\times 10^{-7})$ 
&  BR($B^- \to K ^- K ^0$) & ($\times 10^{-8}$)\\
\hline\hline
  &  without & with &  without & with\\ [-3mm]
& space-like & space-like & space-like & space-like\\[-3mm]
$\epsilon_B$ & penguin & penguin & penguin & penguin\\
\hline
 0.05 & 8.38 & 8.00 & 10.3 & 9.85 \\
\hline
 0.06 & 5.67 & 5.36 & 6.99 & 6.61 \\
\hline
 0.07 & 4.10 & 3.85 & 5.05 & 4.74 \\
\hline
0.08 &  3.11 & 2.90 & 3.83 & 3.57 \\
\hline
0.09 &  2.45 & 2.27 & 3.02 & 2.79  \\ 
\hline
0.10 &  1.99 & 1.83 & 2.45 & 2.25  \\
\hline
\hline
\end{tabular}
\caption{Branching ratios for $B^- \to \pi ^- K ^0$ and
$B^- \to K ^- K ^0$ with space-like penguins
included or not.}
\end{table}

\section{}

We have carried out the analysis on the
effects of the space-like penguin
diagrams within the PQCD method.
General suppressions of the space-like penguins
are found compared to the time-like penguins.
Numerically, the space-like penguins contribute
only several percents in the penguin dominated processes 
$B^- \to \pi ^- K ^0$
and $B^- \to K ^- K ^0$.
We conclude that space-like penguins are less important
in this method.

\section*{}

We thank G. Eilam, M. Gronau, G. Kramer, P. Singer, D. Wyler, and Z.-Z. Xing 
for helpful discussions.
The research of D.-X. Z. was supported in part by Grant 5421-3-96
from the Ministry of Science and the Arts of Israel.

\newpage
\section*{Figure captions}

\noindent 
{\bf 1.} Time-like (a) and space-like (b) penguin diagrams.

\noindent 
{\bf 2.}  Graphs for the time-like penguins in PQCD.
Here the dashed-lines denote the gluons,
and the crosses denote the possible quark-gluon vertices.

\noindent 
{\bf 3.}  Graphs for the space-like penguins in PQCD.

\newpage

\begin{figure}[htb]
\centerline{\epsfxsize8.5in\epsfbox{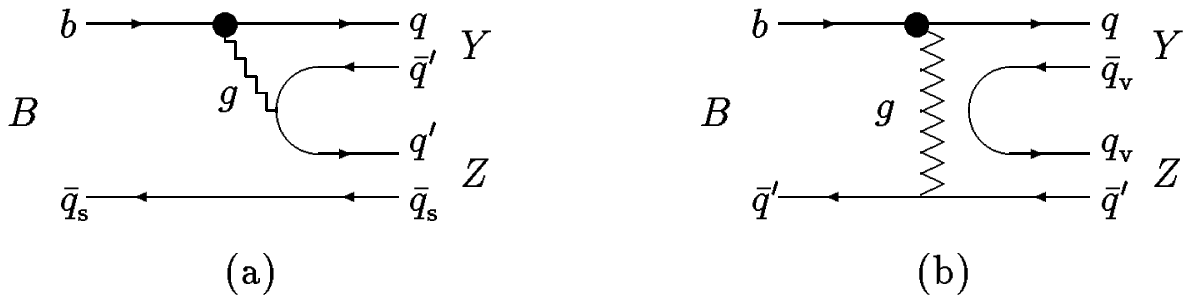}}
\caption{}
\end{figure}

\begin{figure}[htb]
\centerline{\epsfxsize3in\epsfbox{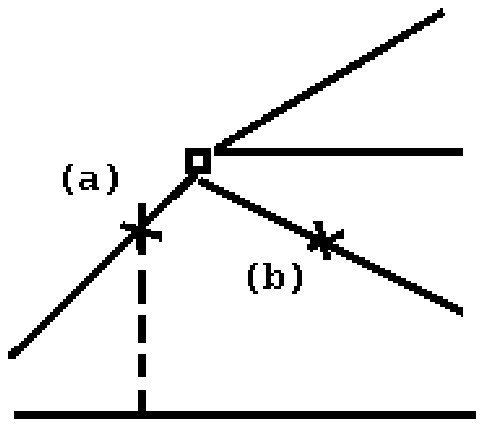}}
\centerline{\epsfxsize3in\epsfbox{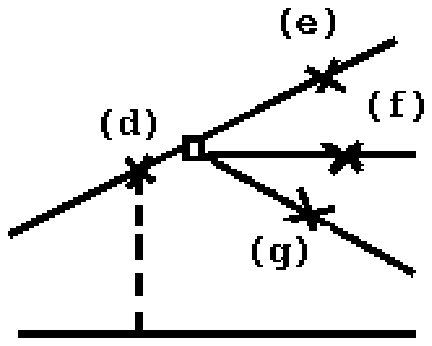}}
\caption{}
\end{figure}

\begin{figure}[htb]
\centerline{\epsfxsize3in\epsfbox{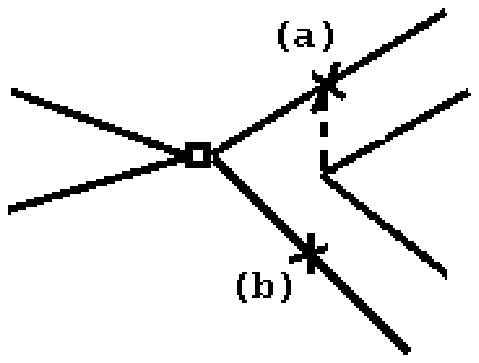}}
\centerline{\epsfxsize3in\epsfbox{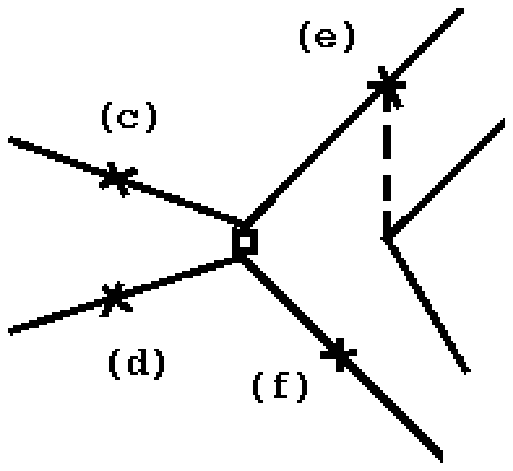}}
\caption{}
\end{figure}
\end{document}